# UI₃ – 5f-electron magnetic van der Waals material

Dávid Hovančík, Marie Kratochvílová, Petr Doležal, Anežka Bendová, Jiří Pospíšil, and Vladimír Sechovský

*Charles University, Faculty of Mathematics and Physics, Department of Condensed Matter Physics, Ke Karlovu 5, 121 16 Prague 2, Czech Republic*

## Abstract

We grew high-quality single crystals of the 5f electron van der Waals compound UI₃ and investigated them by measurements of specific heat and magnetization as functions of temperature and magnetic field. UI₃ behaves as an antiferromagnet with a first-order magnetic phase transition at the Neél temperature $T_N$ = 2.65 K. It is characterized by a sharp symmetric specific-heat peak which is gradually shifted to lower temperatures by increasing magnetic field applied along either the *a*- or *b*-axis. The behavior in magnetic fields reveals orthorhombic magnetocrystalline anisotropy with the hard magnetization direction along the *c*-axis. The 2-K *a*- and *b*-axis magnetization curves exhibit a metamagnetic transition at the critical field $\mu_0 H_c \approx 3$ T and $\approx 1.8$ T, respectively. In higher fields, when the long-range antiferromagnetism is suppressed by a metamagnetic transition, signs of short-range magnetic ordering of antiferromagnetic correlations in the paramagnetic state show up both in specific heat and magnetization. The anomalous S-shape field dependence of *a*-axis magnetization well above $H_c$ can be understood as the crossover from the correlated paramagnetic regime to the high-field polarized paramagnet. The magnetic phase diagrams for the magnetic field applied along the *a*- and *b*-axis, respectively, designed using the mentioned experimental results are presented. We have also found that the UI₃ crystals are easily cleavable which predisposes this material for direct investigation of 5*f* electron magnetism in the 2D limit on exfoliated atomically thin samples.

## 1. Introduction

Magnetic van der Waals (vdW) materials are subjects of intense interest due to their potential use in spintronic and optoelectronic devices [1-6]. The number of materials preserving the magnetic order, especially ferromagnetic, down to the monolayer limit is limited. According to the Mermin-Wagner theorem, in an isotropic Heisenberg system, two-dimensional (2D) long-range ferromagnetism is destroyed at finite temperatures by thermal fluctuations [7]. Strong magnetocrystalline anisotropy promotes stable 2D long-range ferromagnetism which has been experimentally confirmed by observations of a ferromagnetic ground state in atomically thin layers of e.g. CrGeTe₃ [8] or CrI₃ [9]. This property has brought the group of vdW transition metal trihalides MX₃ (M - d-metal, X-halogen atom) to the forefront of scientific interest. The group of trihalides is characterized by a considerable richness of candidates, where the mutual relation between the crystal structure and magnetism can be investigated systematically [10, 11]. Rapid progress has led to the subsequent discovery of ferromagnetism in vanadium triiodide VI₃ [12] in which the unusual magnetocrystalline anisotropy [13] accompanied by two magnetic and structural transitions [14] points to a strong magneto-elastic coupling in these materials [15].

So far, most of the reported vdW 2D magnets are based on $d$-electron transition metals [1, 16]. Recent works suggest that materials based on $f$-electron elements, which possess much stronger spin-orbit interaction and magnetocrystalline anisotropy than the $d$-electron ones, may form another important group of vdW 2D magnets. Theoretical works using density functional theory and the Mont-Carlo method have predicted 2D ferromagnetism with Curie temperature over 100 K in monolayers of a 5f-electron compound UI$_3$ [17]. Even higher Curie temperatures of 160 and 125 K were predicted for UCl$_3$ and UBr$_3$, respectively [18]. The knowledge of the magnetic properties of uranium trihalides is quite limited although they have been well known and widely used as a starting component for organometallic uranium chemistry [19]. The UX$_3$ compounds crystalize in different crystal structures than the MX$_3$ analogs. UF$_3$ is hexagonal (P6$_3$/mcm) [20] UCl$_3$ and UBr$_3$ are hexagonal (C6$_3$/m) [21, 22], and UI$_3$ crystallizes in the orthorhombic PuBr$_3$-type structure (Cmcm) [23], all confirmed by a recent study by Rudel et al. [19]. The historical works report only the basic magnetic properties of UX$_3$, mostly measured on polycrystalline pellets. UF$_3$ orders ferromagnetically at $T_C$ = 1.59 K [24].

The knowledge of UCl$_3$ is controversial. Antiferromagnetic (AFM) ordering below $T_N \approx$ 22-23 K was reported in the work of Hinatsu et al. [25] and Handler et al. [26] while AFM order below $T_N$= 6.5(1) K and an order to order transition at $T_t$ = 3.8(5) K were reported by Schmid et al. [27]. UBr$_3$ has also revealed a more complex magnetic structure with two magnetic transitions at $T_N$ = 5.4(1) K and $T_t$ = 3.0(5) K [27, 28]. A single AFM transition at $T_N$ = 2.61 K and 2.7 K, respectively, was reported for UI$_3$ [29, 30]. To improve the state of knowledge of UI$_3$ physics we have grown the first high-quality single crystals of this compound and measured their specific heat magnetization at temperatures down to 2 K and in a magnetic field up to 14 T applied along the main crystallographic axes. Antiferromagnetic ordering with orthorhombic magnetocrystalline anisotropy has been revealed below $T_N$ = 2.65 K. Unusual magnetization isotherms in the AFM state imply a complicated ground-state magnetic structure. The AFM ordering can be suppressed by magnetic fields comparable in energy scale with $T_N$ but full polarization of U moments can only be achieved in a considerably higher magnetic field. The observed additional corresponding S-shape anomalies on magnetization curves can be understood as the crossover between the correlated paramagnetic regime to the high-field polarized paramagnet.

## 2. Experimental

Although the synthesis of UI$_3$ without the necessity to use a metallic U was proposed [19], we have grown the single crystals by CVT method in evacuated quartz ampoule directly from the stoichiometric ratio (1:3) of elements of purity (U 99.9%, GoodFellow) and (ultra-dried I$_2$ 99.999%, Thermofisher). Several growth regimes were tested to find optimal conditions. The best single crystals in the form of metallic-black plate-like needles were obtained at a reasonable time of three weeks at a temperature gradient 400/500°C. The process was run until the metallic U at the hot end of the tube was fully consumed. The typical length of single crystals was several millimeters however, some exceeded the dimension by a length of over 1 cm. The high quality of the single crystals was verified using the Laue method by patterns with sharp spots recorded on single crystals covered and protected by exfoliation tape (Fig. 1). The Laue method was also used to define the crystallographic directions of the single crystals. The chemical composition of single crystals was verified by scanning electron microscopy (SEM) using a Tescan Mira I LMH system equipped with an energy-dispersive X-ray detector (EDX) Bruker AXS. The analysis revealed a single-phase single crystal of 1:3 composition. Magnetization data were obtained utilizing an MPMS XL 7T and PPMS 14 T both Quantum Design Inc., respectively. The PPMS 14 T apparatus was also used for the specific heat data measurement by relaxation method. All physical quantities were measured with a magnetic field applied along each of the three orthorhombic crystallographic directions. The $UI_3$ single crystals are strongly hygroscopic in air. Therefore, after the growth, the tube with $UI_3$ single crystals was opened in a glovebox with 6N argon protective atmosphere. The storage of single crystals and preparation of all samples for measurements had to be performed under identical protective conditions. The contact of samples with air was reduced to tens of seconds whilst installed inside the instrument under the inert exchange He gas. We were also faced with the problem of sample mounting on measurement holders because of the strong solubility of $UI_3$ in water and ethanol which excluded all common glues. The $UI_3$ was found well resistant to ApiezonN grease which was used as a glue both for magnetization and specific heat measurements. ApiezonN grease is also able to temporarily protect the oiled samples against degradation in air similarly to other trihalides [31]. We extra note that each measurement had to be performed on a new piece of a single crystal. The low stability in the air combined with the extreme cleaving ability of the single crystals did not allow reorientation of one single crystal without damage to all three crystallographic directions concerning the external magnetic field. For a similar reason, we were not able to mount any sample on $^3$He stick of PPMS, and therefore only data down to 2 K could be collected. The mechanical properties and stability of the single crystals also influenced the precise weighing of samples for specific heat and magnetization measurements, with an estimated error 10-20 % in the absolute scale of the plotted data.

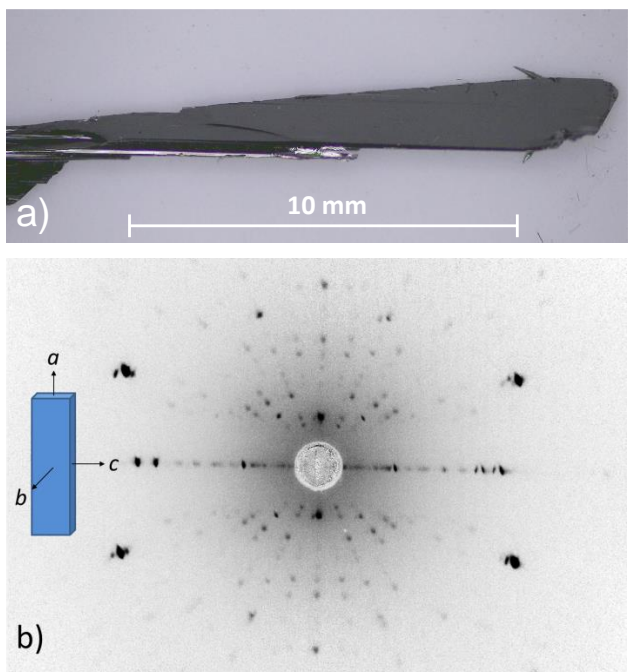

**Fig. 1.** a) The representative $UI_3$ single crystals, b) the Laue pattern recorded on the flat plane of the needle-like single crystal with the illustration of principal crystallographic directions.

## 3. Results

### 3.1. Magnetization

A kink of the temperature dependence of the magnetization (susceptibility) at $\approx 2.6$ K (Fig. 2a) in the magnetic field ($H$) applied along the $a$- and $b$-axis, respectively, can be attributed to the onset of AFM order in $UI_3$. No such feature is seen on the $M(T)$ curve in the $c$-axis field. The 2-K magnetization isotherms recorded in fields applied along corresponding crystallographic directions (Fig. 2b) corroborate the idea of AFM order below 2.6 K. In particular, the S-shape anomaly on the $M(H)$ curve for $H//a$- and $b$-axis most likely reflects a metamagnetic transition at a critical field $\mu_0 H_{c(a)} \approx 3$ T and $\mu_0 H_{c(b)} \approx 1.8$ T, respectively (Fig. 2b). Moreover, the $a$-axis $M(H)$ curve shows a more complex behavior when the rapid growth of magnetization with a characteristic S-shape is still detected above the critical field $H_{c(a)}$. The highest value of the magnetization of $\approx 1.7$ $\mu_B$/f.u. is reached in the maximum magnetic field 14 T. This value is however only about half of the ordered moment calculated for a free $U^{3+}$ or $U^{4+}$ ion.

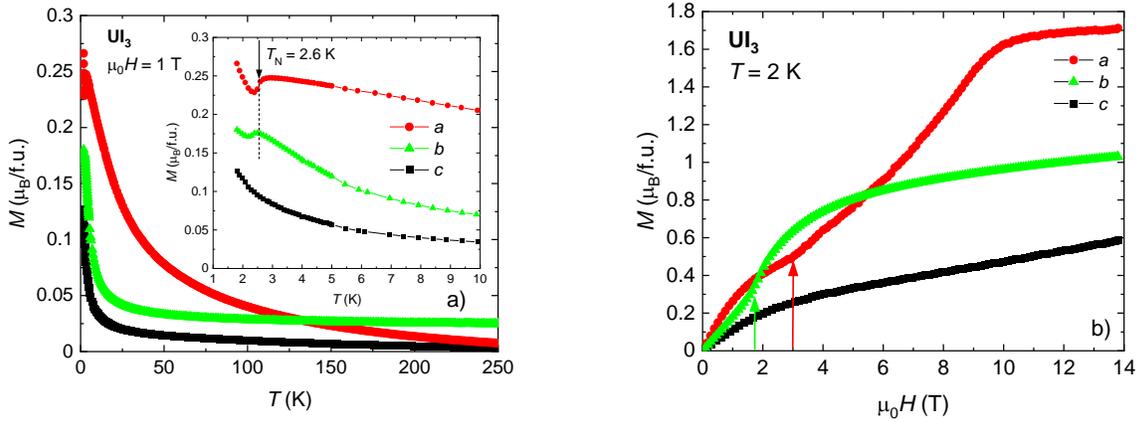

**Fig. 2.** a) The temperature dependence of magnetization and b) the magnetization isotherms at 2 K for all three principal crystallographic directions. The colored arrows in panel b) mark the estimated field of the onset positions of the metamagnetic-like transition for the $a$-axis field $\mu_0 H_{c(a)}$ = 3 T (red) and $\mu_0 H_{c(b)}$ = 1.8 T for the $b$-axis field (green).

Fig. 3 shows the magnetization isotherms for $H \parallel a$ from a wide temperature interval. The original metamagnetic-like transition rapidly vanishes at temperatures approaching $T_N$ (see Fig. 3b). An extra S-shape appears above $H_{c(a)}$ at $\mu_0 H_m \approx 8.5$ T. This S-shape is only weakly affected by increasing temperature and survives above $T_N$ well detectable on the 5-K isotherm. The S-shape completely vanishes at 10 K.

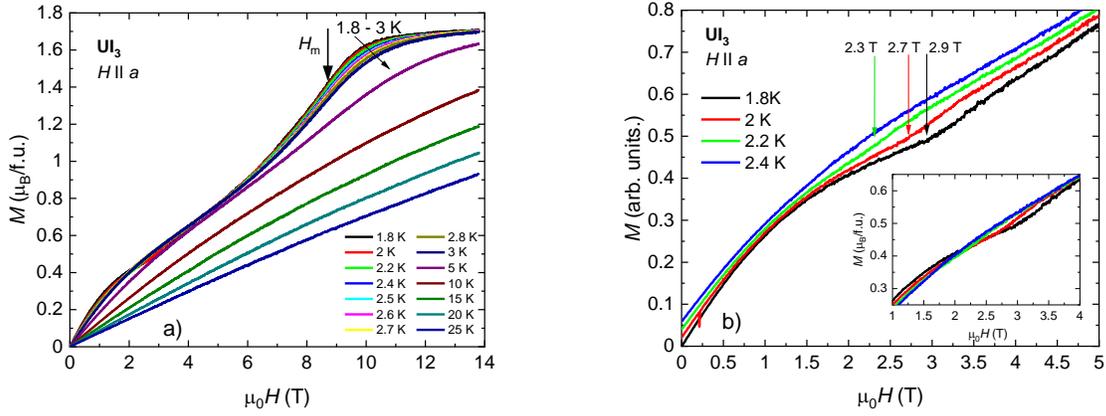

**Fig. 3.** a) The magnetization isotherms of UI$_3$ for *H* ‖ *a* up to the paramagnetic region. The black bold arrow marks the position of the $H_\mathrm{m}$ b) The detail of the low-temperature magnetization isotherms near $H_\mathrm{c(a)}$, the inset shows identical curves at absolute scale.

The magnetization isotherms for *H* ‖ *b* are less complex (Fig. 4). A simple metamagnetic transition was detected already at $\mu_0 H_\mathrm{c(b)} = 1.8$ T (Fig. 4b) but the maximum magnetization reached at 14 T is only $\approx 1.05$ $\mu_\mathrm{B}$/f.u.. The transition is gradually shifted to lower critical field and disappears at temperatures approaching $T_\mathrm{N}$ at a finite rather high magnetic field of 1 T (see the inset of Fig. 4b).

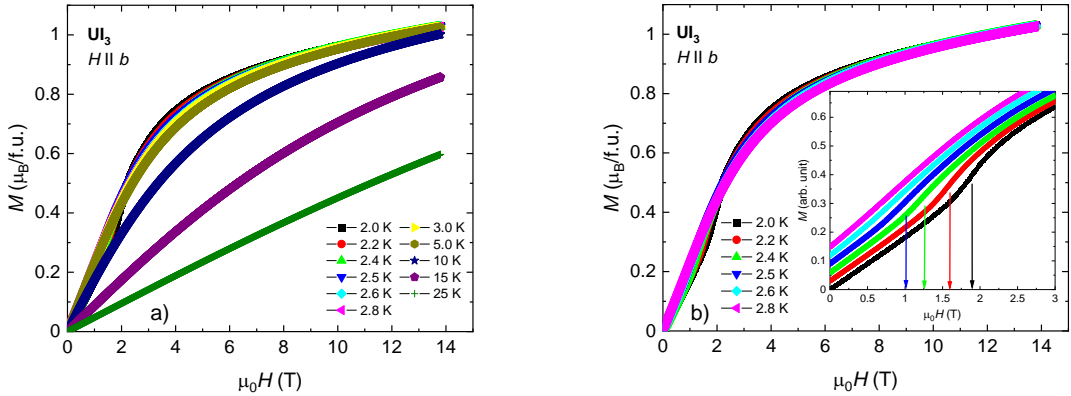

**Fig. 4.** a) The magnetization isotherms of UI$_3$ for *H* ‖ *b* up to the paramagnetic region. b) The detail of the low-temperature magnetization isotherms near $H_\mathrm{c(a)}$, the inset shows identical plots in a relative scale with constant offset for better resolution of them.

The data both for *a* and *b* directions will be used later for the construction of the *H-T* phase diagrams.

*3.2. Specific heat*

The specific heat was measured in the magnetic fields applied along each of the three orthorhombic crystallographic directions. In the zero-field curves, one peak-like anomaly indicating the antiferromagnetic phase transition was detected at 2.7 K (Fig. 5). This agrees well with the observation previously reported for a polycrystalline sample [30]. Upon increasing the magnetic field along the *a*- or *b*-axis, the anomaly rapidly shifts to lower temperatures, which is characteristic of the gradual suppression of AFM order. The effect of the magnetic field is stronger along the *b*-axis direction, where the anomaly disappears below 2 K at the magnetic field of 2 T while along the *a*-axis, the distinct edge of the anomaly is still visible for the magnetic field of 3 T. These field values correspond with the critical fields of metamagnetic transition observed on 2-K magnetization isotherms. On the other hand, the position of the anomaly at $T_N$ remains almost intact by a magnetic field up to 9 T applied along the *c*-axis. This corroborates the idea that the c-axis is to be considered the hard-magnetization direction.

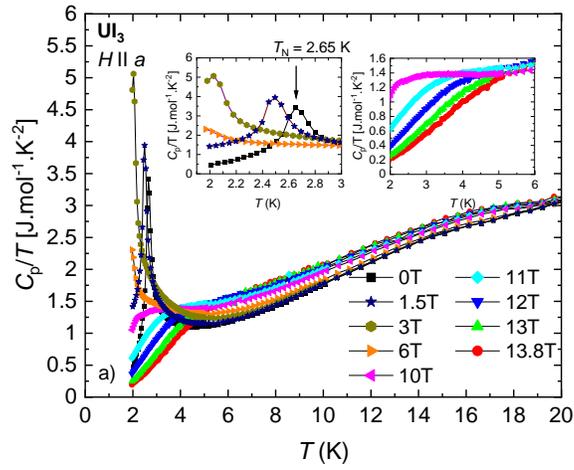

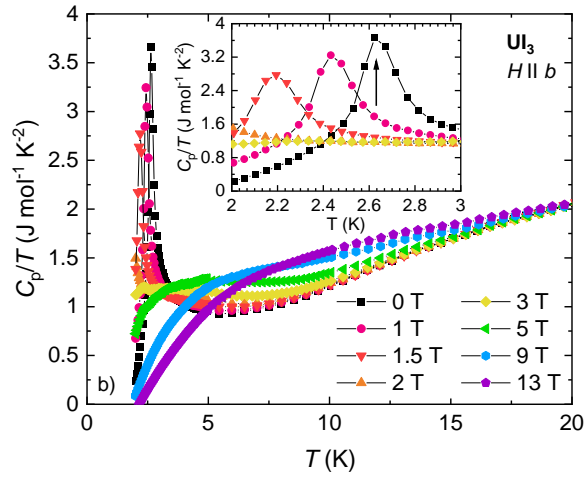

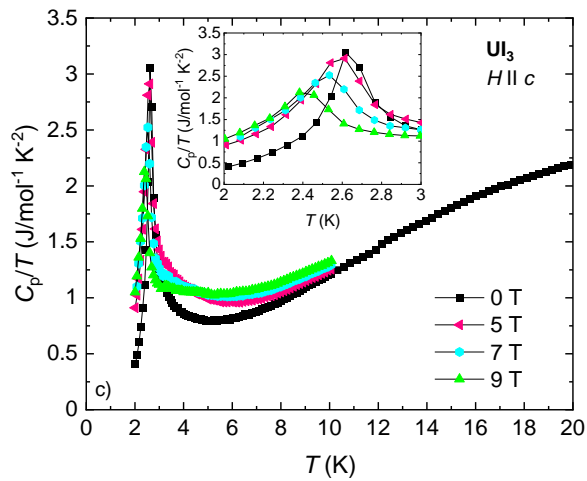

**Fig. 5.** Temperature dependence of the specific heat (plotted as $C_p/T$) of UI$_3$ for magnetic fields applied along all three crystallographic directions a) $H \parallel a$, b) $H \parallel b$, and c) $H \parallel c$. The insets show low-temperature parts with a focus on the evolution of magnetic transition.

## 4. Discussion

In contrast to previous polycrystalline works, our detailed study of the UI$_3$ single crystals has revealed a complex magnetism of this compound. We have found an orthorhombic anisotropy in UI$_3$ where metamagnetic-like transitions were detected both along the $a$- and $b$-axis. However, the axes are not equivalent and finally, the highest (almost saturated) magnetic moment of 1.7 $\mu_B$/f.u. was detected in the field of 14 T applied along the $a$-axis considered as the easy magnetization axis in UI$_3$. The lowest-temperature magnetization behavior measured on UI$_3$ single crystals qualitatively resembles the evolution of metamagnetic transitions in UIrGe in a magnetic field applied along the $b$- and $c$-axis of its orthorhombic structure [32].

The not fully saturated magnetic moment measured on the UI$_3$ single crystal in the $a$-axis magnetic field of 14 T (1.7 $\mu_B$/f.u.) is somewhat smaller than the value of ordered U moment of $\mu_U = 1.98\pm0.05$ $\mu_B$ determined by neutron diffraction experiment [29]. The latter value is rather high, typical for uranium compounds with localized systems. The localized magnetism of UI$_3$ can be also corroborated by crystal structure analysis within the empirical Hill's criterion which suggests a critical interatomic distance $d_{U-U} \approx 3.4$ Å [33] for the boundary of the localized and itinerant limit. The crystal structure of UI$_3$ was solved both by X-ray [19] and neutron diffraction [23], respectively. The coordination number for uranium in UI$_3$ is eight and the coordination polyhedron is a distorted bicapped trigonal prism. The uranium atom is removed from a central position in the prism. The centroid of the prism has coordinates [0.2366, 0, ¼], and deviates from the uranium position [0.2562,0,¼] by 0.275 Å. As six of the U-I bonds are significantly shorter than 3.29 Å, so the U-I bonds have a considerable covalent character [23]. The evaluated $d_{U-U} \approx 4.33$ Å [19] for UI$_3$ locates this compound deep in the localized limit in which a superexchange interaction through I ions will play a dominant role as a mediator of magnetic order [34].

We have fitted the magnetic susceptibility (Fig. 6) by Curie-Weiss and modified Curie-Weiss law and the evaluated parameters are summarized in Tab. I. The effective moment for the easy $a$-axis $\mu_{eff(a)} = 3.42$ $\mu_B$ is very close to that expected for a U$^{3+}$ ion [35] as well as in agreement with the result of Dawson et al. [36]. The metamagnetic transitions along $a$ and $b$-axis both appear near 1/3 value of the saturated magnetization $\mu_{sat}$ which is typical for frustrated AFM systems. The negative value of $\theta_p$ along all three axes agrees with the AFM order of UI$_3$. The frustration factor [37] $f = -\theta_p/T_N \approx 1$ for the $c$-axis while the enhanced value ~5 can be calculated for the $a$- and $b$-axis. First, the values $\theta_p$ oscillate according to the selected section of the paramagnetic region where, particularly, the nonlinear modified Curie-Weiss was applied. Second, the evidence for frustration is considered in materials where the factor $f > 10$ [38], which is not in agreement with our experimental observations.

**Table 1**

The magnetic parameters of UI₃ were evaluated by Curie-Weiss extrapolation of the magnetic susceptibility for the *a*-axis and modified Curie-Weiss for the *b*- and *c*-axis, respectively.

| | $\mu_{\text{eff}}$ () | $\chi_0$ ($10^{-7}$m³/mol) | $\theta_p$ (K) |
|---|---|---|---|
| *a* | 3.42 | - | -11.5 |
| *b* | 1.77 | 1.5 | -14.2 |
| *c* | 1.24 | 0.5 | -3.4 |

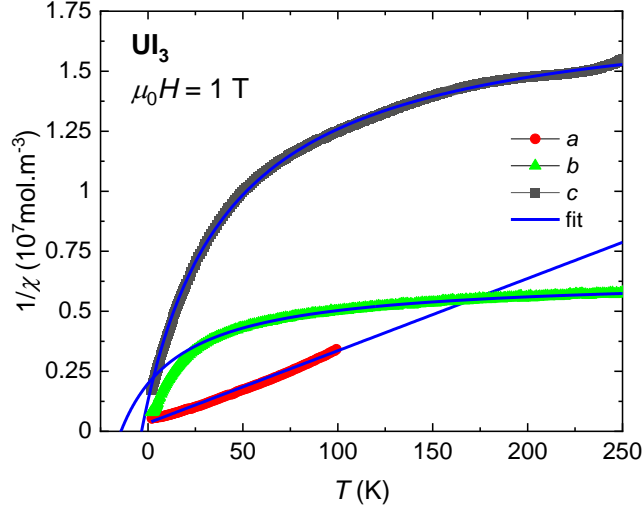

**Fig. 6.** The temperature dependence of inverse magnetic susceptibility with the fits by modified Curie-Weiss and Curie-Weiss law, respectively.

To support the results of the magnetization study, we have analyzed the specific heat data to calculate the magnetic contribution and related magnetic entropy $S_{\text{mag}}$. To determine the magnetic entropy $S_{\text{mag}}$ associated with the AF ordering, the low-temperature phonon and electron part $C_{\text{ph}} + C_e$ has been tentatively separated. The difficulty to extract the phonon contribution $C_{\text{ph}} + C_e$ from UI₃ specific-heat data by a simple Debye formula, which often works for delocalized uranium systems with low ordering temperature [39-43] can be connected to additional contribution $C_{\text{add}}$ due to short-range ordering or AFM correlations in the paramagnetic state. Since non-magnetic analog data are not available, a general polynomial function has been used. We have fitted zero-field data for all three axes with a negligible difference –see the example for the *a*-axis in Fig. 7. The average value of the overall three crystallographic directions is $S_{\text{mag}} = 0.52$Rln2 which is the signature of the strong localization of the U electrons, however, $S_{\text{mag}} \approx$ Rln2 is the expectable value when $\mu_{\text{eff(a)}} = 3.42$ $\mu_B$ is very close to for U³⁺ ion. A similar analysis was performed for the localized ferromagnet UF₃ ($T_C = 1.59$ K) with $S_{\text{mag}} =$ Rln2 [24] thanks to the ⁴I₉/₂ configuration of U³⁺ splitting by the crystal field resulting in a doublet ground state well separated (~142 K) from the first excited doublet, confirmed spectroscopically [44]. The inelastic neutron scattering study of UBr₃ ($\mu_{\text{eff}} = 3.57$ $\mu_B$ by Curie-Weiss law) has also shown a strong crystal-field effect and splitting of the ⁴I₉/₂ ground state multiplet with the first excited level 4.1 meV [28]. We could only speculate based on the reduced $S_{\text{mag}}$ value a low-lying first excited multiplet in UI₃, which is also projected to strong curvature of the $1/\chi$ data. However such a scenario is not in agreement with observed three ground-state crystal field transitions at 15, 25, and 38 meV in $UI_3$ observed by inelastic neutron diffraction [34], which is comparable with $UF_3$ and $UBr_3$ crystal field splitting schemes.

The characteristic feature of the AFM transition in specific heat data in $UI_3$ is a peak-like transition signaling the first-order transition [34]. A similar effect of the so-called "Missing entropy" in $UI_3$ [45] was reported in antiferromagnet $UO_2$ [46] in which the AFM transition is also of the first-order type explained by the Blume model [47].

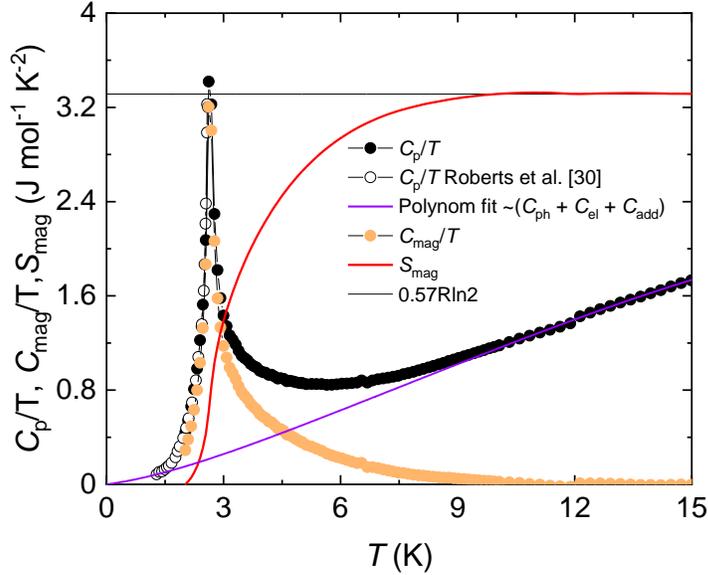

**Fig. 7.** The temperature dependence of zero-field specific heat with extracted magnetic contribution $C_{mag}$ and calculated $S_{mag}$. The sum of the phonon $C_{ph}$, electron $C_{el}$, and $C_{add}$ contributions was estimated by a polynomial function. The results of Roberts and Murray [30], in good agreement with our results, are involved to extend the temperature scale for better evaluation of the data.

A scenario explaining the "Missing entropy" in the $UI_3$ case is a short-range order at temperatures well above $T_N$ [45]. It motivated us to construct the $H$-$T$ phase diagram of $UI_3$ for the magnetic field applied along the $a$- and $b$-axis, respectively (Fig. 8). In the $a$-axis case, an extra $S$-shape around $H_m > 2H_{c(a)}$ has been observed at sufficiently low temperatures (Fig. 3a and 8a). The $S$-shape is gradually smearing out with increasing temperature. It is still well detectable at 5 K and fully vanishes at a temperature of 10 K. We note that the effect of the $S$-shape was detectable exclusively in fields applied along the $a$-axis. The S-shape of the magnetization curve above a metamagnetic transition is usually explained as the crossover from the correlated paramagnetic regime CPM (paramagnet with AFM correlations or AFM short-range order) to the polarized paramagnetic regime (PPM) [48, 49].

The peak-like anomaly at the AFM transition in specific heat data responds to the external magnetic field in agreement with the gradual vanishing of the metamagnetic transition. In ordinary antiferromagnets, a paramagnetic-like response of the temperature dependence of specific heat is detected for fields $H > H_c$ [50, 51]. Broad knees have been observed in specific heat data both for the $a$- and $b$-axis in the curves recorded for magnetic fields $H > H_{c(a,b)}$. The temperature dependence of specific heat recorded in the highest available magnetic field 13.8 T still shows a pronounced anomaly (Figs. 5a and 5b) and a significantly higher magnetic field would be necessary to fully vanish it. Moreover, the 13.8 T curves merge with the zero-field data at a rather high temperature of 20 K far above $T_N$.

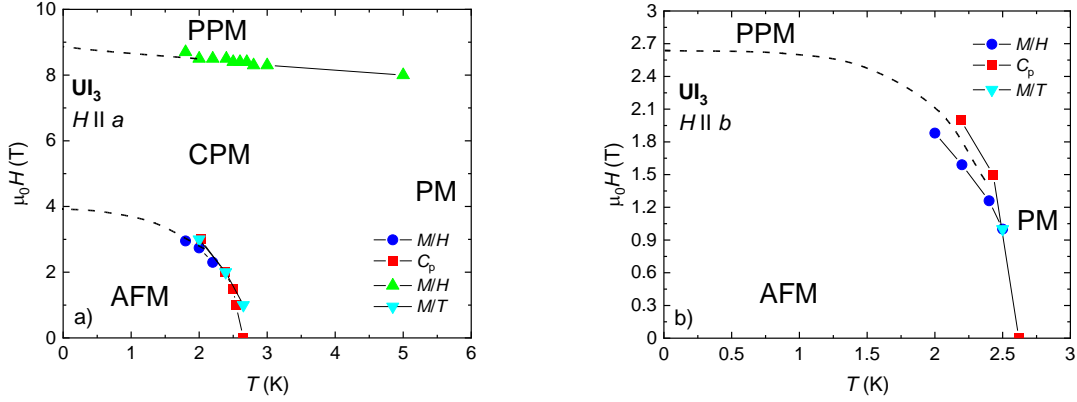

**Fig. 8.** Magnetic phase diagrams of UI$_3$ for $H$ parallel to the $a$-axis a) and $b$-axis b).

In the frame of the propagation vector $\vec{k} = (1/2, 1/2, 1/2)$ of the AFM structure determined by neutron-diffraction experiment [29], we can only deduce a commensurate AFM structure with magnetic moment direction located within the $ab$-plane, with the dominant component on the $a$-axis.

## 5. Conclusions

We have successfully prepared large UI$_3$ single crystals and studied them by measuring the specific heat and magnetization methods as functions of temperature and magnetic field applied in each of the main crystallographic directions. The results point to a complex AFM ordering with the first-order transition at $T_N = 2.6$. The AFM state shows the orthorhombic magnetocrystalline anisotropy with the easiest(hardest) magnetization direction along the $a(c)$-axis. The Curie-Weiss behavior of paramagnetic susceptibility data in fields applied along the easy axis corroborates a localized 5f-electron magnetism with the effective moment consistent with the calculated value for a free U$^{3+}$ ion. The determined magnetic entropy at Néel temperature $S_{mag} \sim \frac{1}{2}$ Rln2 is much lower than expected. The analysis of the $a$-axis magnetization and specific heat data in magnetic fields well above the metamagnetic transition points to a possible short-range AFM order in the paramagnetic state. Relevant microscopic (e.g. neutron diffraction) experiment on sufficiently large and stable single crystals is highly desirable to figure out this aspect of UI$_3$ magnetism. We also observed that the UI$_3$ single crystals are well and simply cleavable that predestines this material for the investigation of the 5$f$ electron magnetism in 2D limit. However, the poor stability of currently available crystals in a normal atmosphere limits the complicated sample preparation for some experiments.

**Acknowledgments**

This work is part of the research program GACR 21-06083S which is financed by the Czech Science Foundation. This project was also supported by OP VVV project MATFUN under Grant No. CZ.02.1.01/0.0/0.0/15_003/0000487. Experiments were performed in MGML (https://mgml.eu/), which is supported by the program of Czech Research Infrastructures (project no. LM2018096). We also acknowledge the technical assistance of Dr. M. Míšek from MGML. The authors are alo indebted to Dr. Ross Colman for critical reading and correcting of the manuscript.

References

[1]   X. Jiang, Q.X. Liu, J.P. Xing, N.S. Liu, Y. Guo, Z.F. Liu, J.J. Zhao, Applied Physics Reviews, 8 (2021) 031305.

[2]   W. Zhang, P.K.J. Wong, R. Zhu, A.T.S. Wee, InfoMat, 1 (2019) 479-495.

[3]   K.S. Burch, D. Mandrus, J.G. Park, Nature, 563 (2018) 47.

[4]   D. Zhong, K.L. Seyler, X.Y. Linpeng, R. Cheng, N. Sivadas, B. Huang, E. Schmidgall, T. Taniguchi, K. Watanabe, M.A. McGuire, W. Yao, D. Xiao, K.M.C. Fu, X.D. Xu, Science Advances, 3 (2017).

[5]   J.-G. Park, Journal of Physics: Condensed Matter, 28 (2016) 301001.

[6]   P. Ajayan, P. Kim, K. Banerjee, Physics Today, 69 (2016) 39-44.

[7]   N.D. Mermin, H. Wagner, Physical Review Letters, 17 (1966) 1133-&.

[8]   C. Gong, L. Li, Z. Li, H. Ji, A. Stern, Y. Xia, T. Cao, W. Bao, C. Wang, Y. Wang, Z.Q. Qiu, R.J. Cava, S.G. Louie, J. Xia, X. Zhang, Nature, 546 (2017) 265–269.

[9]   B. Huang, G. Clark, E. Navarro-Moratalla, D.R. Klein, R. Cheng, K.L. Seyler, D. Zhong, E. Schmidgall, M.A. McGuire, D.H. Cobden, W. Yao, D. Xiao, P. Jarillo-Herrero, X.D. Xu, Nature, 546 (2017) 270-+.

[10]  M.A. McGuire, Crystals, 7 (2017) 121.

[11]  M. Kratochvílová, P. Doležal, D. Hovančík, J. Pospíšil, A. Bendová, M. Dušek, V. Holý, V. Sechovský, Journal of Physics: Condensed Matter, 34 (2022) 294007.

[12]  T. Kong, K. Stolze, E.I. Timmons, J. Tao, D.R. Ni, S. Guo, Z. Yang, R. Prozorov, R.J. Cava, Advanced Materials, 31 (2019) 1808074.

[13]  A. Koriki, M. Míšek, J. Pospíšil, M. Kratochvílová, K. Carva, J. Prokleška, P. Doležal, J. Kaštil, S. Son, J.-G. Park, V. Sechovský, Physical Review B, 103 (2021) 174401.

[14]  P. Doležal, M. Kratochvílová, V. Holý, P. Čermák, V. Sechovský, M. Dušek, M. Míšek, T. Chakraborty, Y. Noda, S. Son, J.G. Park, Physical Review Materials, 3 (2019) 121401(R).

[15]  J. Valenta, M. Kratochvílová, M. Míšek, K. Carva, J. Kaštil, P. Doležal, P. Opletal, P. Čermák, P. Proschek, K. Uhlířová, J. Prchal, M.J. Coak, S. Son, J.-G. Park, V. Sechovský, Physical Review Materials, 103 (2021) 054424.

[16]  M.C. Wang, C.C. Huang, C.H. Cheung, C.Y. Chen, S.G. Tan, T.W. Huang, Y. Zhao, Y.F. Zhao, G. Wu, Y.P. Feng, H.C. Wu, C.R. Chang, Annalen Der Physik, 532 (2020) 1900452.

[17]  S.J. Li, Z.T. Wang, M. Zhou, F.W. Zheng, X.H. Shao, P. Zhang, Journal of Physics D-Applied Physics, 53 (2020) 185301.

[18]    S.J. Li, M. Zhou, X.H. Wang, F.W. Zheng, X.H. Shao, P. Zhang, Physics Letters A, 394 (2021) 127078.

[19]    S.S. Rudel, H.L. Deubner, B. Scheibe, M. Conrad, F. Kraus, Zeitschrift Fur Anorganische Und Allgemeine Chemie, 644 (2018) 323-329.

[20]    E. Staritzky, R. Douglass, Analytical Chemistry, 28 (1956) 1056-1057.

[21]    W.H. Zachariasen, The Journal of Chemical Physics, 16 (1948) 254-254.

[22]    W.H. Zachariasen, The Crystal Structure of Trichlorides, Tribromides and Trihydroxides of Uranium and of Rare Earth Elements in:  U.S. Atomic Energy Commission. Technical Information Service, Oak Ridge, Tenn., 1948, pp. AECD-2091.

[23]    J.H. Levy, J.C. Taylor, P.W. Wilson, Acta Crystallographica, B31 (1975) 880.

[24]    O. Beneš, J.C. Griveau, E. Colineau, D. Sedmidubský, R.J.M. Konings, Inorganic Chemistry, 50 (2011) 10102-10106.

[25]    Y. Hinatsu, C. Miyake, S. Imoto, Journal of Nuclear Science and Technology, 17 (1980) 929-934.

[26]    P. Handler, C.A. Hutchison, The Journal of Chemical Physics, 25 (1956) 1210-1213.

[27]    B. Schmidt, A. Murasik, P. Fischer, A. Furrer, B. Kanellakopulos, Journal of Physics: Condensed Matter, 2 (1990) 3369-3380.

[28]    R. Lyzwa, P. Erdös, Physical Review B, 36 (1987) 8570-8573.

[29]    P. Fischer, A. Murasik, W. Szczepaniak, Helvetica Physica Acta, 53 (1980) 577-577.

[30]    L.D. Roberts, R.B. Murray, Physical Review, 100 (1955) 650-654.

[31]    M. Kratochvílová, K. Uhlířová, M. Míšek, V. Holý, J. Zázvorka, M. Veis, J. Pospíšil, S. Son, J.G. Park, V. Sechovský, Mater Chem Phys, 278 (2022) 125590.

[32]    S. Yoshii, A.V. Andreev, E. Brück, J.C.P. Klaasse, K. Prokeš, F.R. de Boer, M. Hagiwara, K. Kindo, V. Sechovský, Journal of Physics: Conference Series, 51 (2006) 151.

[33]    H.H.Hill, Plutonium 1970 and Other Actinides ed. W. N.Miner, (American Institute of Mining, Metalurgical, and Metalurgical Engeneers, New York), p. 2, 1970.

[34]    A. Murasik, P. Fischer, A. Furrer, B. Schmid, W. Szczepaniak, Journal of the Less-Common Metals, 121 (1986) 151-155.

[35]    M. Vališka, J. Pospíšil, A. Stunault, Y. Takeda, B. Gillon, Y. Haga, K. Prokeš, M. M. Abd-Elmeguid, G. Nenert, T. Okane, H. Yamagami, L. Chapon, A. Gukasov, A. Cousson, E. Yamamoto, V. Sechovský, Journal of the Physical Society of Japan, 84 (2015) 084707.

[36]    J.K. Dawson, Journal of the Chemical Society, (1951) 429-431.

[37]    J. Pospisil, G. Nenert, S. Miyashita, H. Kitazawa, Y. Skourski, M. Divis, J. Prokleska, V. Sechovsky, Physical Review B, 87 (2013) 214405.

[38]    A.P. Ramirez, Annual Review of Materials Science, 24 (1994) 453-480.

[39]    J. Pospíšil, Y. Haga, S. Kambe, Y. Tokunaga, N. Tateiwa, D. Aoki, F. Honda, A. Nakamura, Y. Homma, E. Yamamoto, T. Yamamura, Physical Review B, 95 (2017) 155138.

[40]    K. Prokeš, T. Tahara, T. Fujita, H. Goshima, T. Takabatake, M. Mihalik, A.A. Menovsky, S. Fukuda, J. Sakurai, Physical Review B, 60 (1999) 9532-9538.

[41]    J. Pospisil, Y. Haga, A. Miyake, S. Kambe, Y. Tokunaga, M. Tokunaga, E. Yamamoto, P. Proschek, J. Volny, V. Sechovsky, Physical Review B, 102 (2020) 024442.

[42]    F. Hardy, D. Aoki, C. Meingast, P. Schweiss, P. Burger, H. Von Löhneysen, J. Flouquet, Physical Review B, 83 (2011) 195107.

[43]    K. Prokes, T. Tahara, Y. Echizen, T. Takabatake, T. Fujita, I.H. Hagmusa, J.C.P. Klaasse, E. Brück, F.R. de Boer, M. Divis, V. Sechovsky, Physica B: Condensed Matter, 311 (2002) 220-232.

[44]    M. Karbowiak, J. Drozdzynski, Chem. Phys., 340 (2007) 187-196.

[45]    S.I. Parks, W.G. Moulton, Physical Review, 173 (1968) 333-337.

[46]    B.C. Frazer, G. Shirane, D.E. Cox, C.E. Olsen, Physical Review, 140 (1965) A1448-A1452.

[47]    M. Blume, Physical Review, 141 (1966) 517-524.

[48]    J. Pospíšil, Y. Haga, Y. Kohama, A. Miyake, K. Shinsaku, T. Naoyuki, M. Vališka, P. Proschek, J. Prokleška, V. Sechovsky, M. Tokunaga, K. Kindo, A. Matso, E. Yamamoto, Physical Review B, 98 (2018) 014430.

[49]    W. Knafo, R. Settai, D. Braithwaite, S. Kurahashi, D. Aoki, J. Flouquet, Physical Review B, 95 (2017) 014411.

[50]    J. Pospíšil, M. Míšek, M. Diviš, M. Dušek, F.R. de Boer, L. Havela, J. Custers, Journal of Alloys and Compounds, 823 (2020) 153485.

[51]    F. Honda, J. Valenta, J. Prokleška, J. Pospíšil, P. Proschek, J. Prchal, V. Sechovský, Physical Review B, 100 (2019) 014401.